\documentclass[english,prd,twocolumn,nofootinbib,floatfix]{revtex4}
\usepackage[T1]{fontenc}
\usepackage[latin9]{inputenc}
\usepackage{amsmath}
\usepackage{amssymb}
\usepackage{graphicx,epsfig}

\makeatletter
\@ifundefined{textcolor}{}
{%
 \definecolor{BLACK}{gray}{0}
 \definecolor{WHITE}{gray}{1}
 \definecolor{RED}{rgb}{1,0,0}
 \definecolor{GREEN}{rgb}{0,1,0}
 \definecolor{BLUE}{rgb}{0,0,1}
 \definecolor{CYAN}{cmyk}{1,0,0,0}
 \definecolor{MAGENTA}{cmyk}{0,1,0,0}
 \definecolor{YELLOW}{cmyk}{0,0,1,0}
 }

\usepackage{amsfonts}
\providecommand{\U}[1]{\protect\rule{.1in}{.1in}}%EndMSIPreambleData
\newcommand{\BOX}{\hbox {$\sqcap$ \kern -1em $\sqcup$}}\newcommand{\be}{\begin{equation}}\newcommand{\ee}{\end{equation}}\newcommand{\ba}{\begin{eqnarray}}\newcommand{\ea}{\end{eqnarray}}\newcommand{\ban}{\begin{eqnarray*}}\newcommand{\bea}{\begin{eqnarray}}\newcommand{\eea}{\end{eqnarray}}\newcommand{\ean}{\end{eqnarray*}}\newcommand{\barr}{\begin{array}}\newcommand{\earr}{\end{array}}

\makeatother

\usepackage{babel}

\makeatother

\usepackage{babel}

\makeatother

\usepackage{babel}

\makeatother

\usepackage{babel}

\begin{document}
CERN-PH-TH/2010-302
\par KCL-PH-TH/2010-35
\vspace*{1cm}

\title{On the Interpretation of Gravitational Corrections to Gauge Couplings}

\author{John Ellis and Nick E.\ Mavromatos}
\affiliation{CERN, Theory Division, CH-1211  Geneva 23, Switzerland; \\ King's College London, Department of Physics, Strand, London WC2R~2LS, UK}

\begin{abstract}
Several recent papers discuss gravitational corrections to gauge couplings that depend
quadratically on the energy. In the framework of the background-field approach,
these correspond in general to adding to the effective action terms quadratic in the field
strength but with higher-order space-time derivatives. We observe that such terms
can be removed by appropriate local field redefinitions, and do not contribute to
physical scattering-matrix elements. We illustrate this observation in the context of open
string theory, where the effective action includes, among other terms, the well-known
Born-Infeld form of non-linear electrodynamics. We conclude that the quadratically
energy-dependent gravitational corrections are \emph{not} physical in the sense of
contributing to the running of a physically-measurable gauge coupling, or of unifying
couplings as in string theory.

\end{abstract}

\maketitle

%\date{\today}

Much discussion has recently been stimulated by~\cite{wilczek}, in which the contribution of graviton exchange to the one-loop three gluon vertex in non-Abelian gauge theory was calculated. It was argued in~\cite{wilczek} that there are non-trivial contributions to the one-loop renormalization-group $\beta$ function of the gauge coupling $g$ that exhibit a quadratic dependence on the energy $E$, which are suppressed by the square of the Planck scale $M_P$, the characteristic scale of gravitational interactions:
\begin{equation}\label{quadrbeta}
\beta (g, E) = \frac{d\, g}{d\,{\rm ln}E} = \frac{b_0}{4\pi^2}\,g^3 - 3\frac{16\pi^2}{(4\pi)^2}\frac{E^2}{M_P^2}\,g ~.
\end{equation}
The first term is the standard one appearing in the absence of gravitation, corresponding to a \emph{logarithmic}
dependence on the energy $E$ of the renormalized gauge coupling $g$,
whose coefficient has the same value as in a pure gauge theory, given that graviton does not carry any gauge charge.
The second term with a quadratic dependence on $E$ is due to the gravitational contribution,
which was argued in \cite{wilczek}
as making the running coupling asymptotically free, even if $b_0 > 0$, and thereby avoiding a Landau pole in QED. This observation would also have profound implications for the unification with gravity with the gauge interactions, since it would cause their couplings to drop very rapidly at scales above the expected grand unification
scale $\sim 10^{16}$ GeV~\cite{wilczek}.

It was argued in~\cite{gauge} that these results depend on the gauge-fixing parameter, which would imply that the gravitational contributions are not physical. Additionally, it was commented in~\cite{absent} that, because of their
quadratically-divergent nature, the corrections in~\cite{wilczek} would vanish in dimensional
regularization~\cite{absent}, an observation which was consistent with the cancellation of
quadratic divergences in a momentum-space cutoff approach~\cite{various}. However, other regularization approaches~\cite{other}
have produced various non-trivial results for gravitational corrections to the running of the gauge couplings.

A covariant approach to the one-loop effective action was taken recently in~\cite{toms},
based on a gauge-invariant heat-kernel regularization, and it was claimed in~\cite{toms} that the
quadratically energy-dependent one-loop gravitational contributions to the effective action of QED do not cancel.
Ref.~ \cite{toms} went beyond previous calculations by including in the gravitational sector
a positive cosmological constant $\Lambda > 0$. The result of this one-loop analysis using the background-field
method and expanding the gravitational corrections around a Minkowski background, was to produce a
gauge-invariant QED effective action with quadratic and logarithmic dependences on energy, of the form:
\begin{eqnarray}\label{greff}
&& {\rm Div}\Gamma^{(1)} = \nonumber \\
 && \left(\frac{e^2}{48\,\pi^2}{\rm ln}E_c^2 - \frac{\kappa^2 E_c^2 }{128\,\pi^2} -
 \frac{\kappa^2 \, \Lambda}{256\,\pi^2}{\rm ln}E_c^2  \right) \int d^4 x F_{\mu\nu} F^{\mu\nu}~, \nonumber \\
\end{eqnarray}
where $\kappa^2 \equiv 32\pi G_N \equiv 16\pi^2/M_P^2$ is the (dimensionful) gravitational coupling
constant and $G_N$ is the Newton constant. Here, the space-time indices are raised and lowered
with respect the Minkowski background metric, the quantity $F_{\mu\nu}$ denotes the background
gauge field strength, $e$ is the electron charge that plays the r\^ole of the coupling in QED,
and $E_c$ is an energy cutoff which, in the approach of~\cite{toms}, is taken to be the inverse of a
proper time cutoff $E_c \equiv 1/\tau_c$. The first term in (\ref{greff}) is the conventional logarithmic
renormalization of the QED coupling that would lead to a Landau pole, the quadratic energy
dependence of the second term in (\ref{greff}) reflects that of the second term in (\ref{quadrbeta}),
and the third term in (\ref{greff}) indicates that
the cosmological constant $\Lambda > 0$ is associated with a logarithmic energy dependence.

These energy-dependent terms were absorbed in~\cite{toms} in a redefinition of a renormalized
electric charge, $e_R$, defined through $\frac{1}{4e_R^2} \int d^4x F_{\mu\nu}F^{\mu\nu} $,
whose one-loop renormalization is described by a $\beta$-function of the following form (identifying the
cutoff $E_c$ with $E$):
\begin{equation}\label{qedbeta}
\beta(E, e_R) = \frac{e_R^3}{12\pi^2} - \frac{\kappa^2}{32\pi^2}\left(E^2 + \frac{3}{2}\Lambda\right) e_R~.
\end{equation}
The first term is the standard one-loop coefficient of QED, while the other terms in parenthesis denote the gravitational contributions. These render asymptotically free the corresponding QED running coupling,
as was originally suggested in~\cite{wilczek}, and similar results characterize non-Abelian gauge theories.
Note, however, that the coefficients of the terms quadratic in $E$ are different in the two calculations
(\ref{quadrbeta}, \ref{qedbeta}). Moreover, another recent analysis~\cite{recent}, using a different gauge-invariant approach to gravitation
employing the Vilkovisky-De-Witt formalism, also finds quadratic corrections, but with the opposite sign,
leading to a conclusion entirely different from~\cite{wilczek,toms} for the gravitational
contributions to the running gauge couplings, namely that they do \emph{not} lead to asymptotic freedom.

The above confusing and contradictory results raise the issue whether the claimed gravitational
corrections to gauge couplings are physical, and specifically the question whether
the quadratic energy dependence of gravitational corrections to terms in an effective action actually
signal the appearance of a running coupling in \emph{physical processes}.

Our starting-point in this letter is to recall the scope and importance of local field redefinitions and the invariance of on-shell scattering amplitudes ($S$-matrix elements)
under such redefinitions, as shown in the equivalence theorem~\cite{equiv}.
This is commonly used in string theory to isolate unambiguous terms in the low-energy effective actions
characterizing strings propagating in non-trivial backgrounds~\cite{tseytlin}. We use the equivalence
theorem to argue that energy-dependent modifications of the gauge coupling such as those discussed
above do not affect $S$-matrix elements and are not relevant for the unification of gauge interactions
with gravity in string theory.

The equivalence theorem can be framed as follows: consider the generating functional of the (tree-level) field-theory $S$-matrix , $\Gamma [\phi_{\rm in} ] = W[J_{\rm in}]$, where the $\phi_{\rm in}$ are ``in'' state fields,
the $J_{\rm in}$ are the corresponding sources, $J_{\rm in} = \phi_{\rm in} \Box $, where $\Box$ is the covariant
d'Alembertian, and the effective action is defined through an appropriate Legendre transform:
\begin{eqnarray}\label{legendre}
&& \Gamma [\phi_{\rm in}] = \Gamma_0 [\phi_0] - \int \phi_{\rm in} \Box \phi_0, \nonumber \\
&& \Gamma_0 = \frac{1}{2}\int \phi \Box \phi + \mathcal{V}(\phi)~,
\end{eqnarray}
where $\phi_0 $ is the perturbative solution of the equations $\delta \Gamma_0 /\delta \phi = 0$, the
equations of motion. The fields
$\phi$ above are generic, and include gauge fields.

The equivalence theorem~\cite{equiv} asserts that if one performs
the redefinition
\begin{equation}
\phi \rightarrow \phi = \tilde{\phi} + \mathcal{T}(\tilde{\phi})~,
\end{equation}
where $\mathcal{T}(\phi)$ is a local, gauge invariant combination of $\phi$ and its derivatives
that does not influence the mass-shell condition, and the correlators of
$\mathcal{T}$ with itself and/or $\phi$ do not have massless poles, then the generating functional
for the $S$-matrix in the transformed theory with the action $\Gamma_0 [\tilde{\phi}]=\Gamma_0 [\phi]$
is the same  as in the original theory.

We argue that the equivalence theorem applies to the claimed gravitational corrections to gauge couplings,
and hence they have no physical effects on on-shell scattering processes. This also explains the discrepancies
described above and the apparent dependences on the gauge-fixing parameter and the regularization scheme.

From the point of view of an effective action, the $E^2$-dependent term in (\ref{greff}) corresponds to a
higher-derivative term of the form:
\begin{equation}\label{quadr}
\frac{b}{M_P^2} \int d^4 x F_{\mu\nu} \Box F^{\mu\nu}~,
\end{equation}
where $b$ is a dimensionless numerical constant. This is the only independent higher-derivative combination
that is quadratic in the field strength and in space-time derivatives. Thanks to the cyclic permutation identity
$\partial_{\mu} F_{\nu\rho} + \partial_{\nu} F_{\rho\mu} + \partial_{\rho} F_{\mu\nu} = 0$,
terms of the form $(\nabla_\rho F_{\nu\mu} )^2$ can easily be cast in the form (\ref{quadr}).

It is straightforward to see that the coefficient of a term such as (\ref{quadr}) can be changed by the following local field redefinition of the gauge potential, $A_\mu$, which respects the criteria of the equivalence theorem outlined above:
\begin{equation}\label{gaugered}
A_\mu \rightarrow A_\mu = \tilde{A}_\mu + \frac{c}{M_P^2}\nabla^\nu F_{\nu\mu}~,
\end{equation}
where $c$ is an arbitrary numerical constant and $\nabla_\mu$ denotes a gravitationally-covariant derivative.
In fact, all photon propagator corrections can be removed by such local redefinitions, and all terms with an arbitrary number of derivatives that are bilinear in the gauge fields $A_{\mu}$, such as  $(\nabla_\rho \dots \nabla_\lambda F_{\alpha\beta})^2$ etc. are ambiguous~\cite{tseytlin} in the sense that on-shell scattering amplitudes are insensitive to their presence~\cite{fermions}.

This observation can be extended to terms of higher orders in the gauge theory effective action.
Specifically, the most general local, gauge-invariant effective action in the pure gauge sector, including
gauge- and gravitationally-covariant space-time derivatives, is known to be~\cite{tseytlin}
\begin{eqnarray}\label{efa}
&&\mathcal{L} = F_{\mu\nu}^2 + 2 \tilde{\kappa}^2 [ a_1 F_{\mu\nu} F^{\nu\rho} F_{\rho}^\mu + a_2 (\mathcal{D}^\rho F_{\rho\mu})^2 ] + \nonumber \\
&& (2\tilde{\kappa}^2)^2 \left[ a_3 F_{\mu\nu}F^{\mu\lambda} F^{\rho\nu}F_{\rho\lambda} + a_4 F_{\mu\nu}F^{\mu\lambda} F^{\rho\lambda}F_{\rho\nu} + \right. \nonumber \\
&& \left.  a_5 (F_{\mu\nu}F^{\mu\nu})^2 + a_6 F_{\mu\nu} F_{\lambda\rho}F^{\mu\nu} F^{\lambda\rho} + \right.\nonumber \\
&& \left. a_7 F_{\mu\nu} \mathcal{D}^\lambda F^{\mu\nu} \mathcal{D}^\rho F_{\rho\lambda} + a_8 \mathcal{D}^\lambda F_{\lambda \mu} \mathcal{D}^\rho F_{\rho\nu} F^{\mu\nu} + \right. \nonumber \\
&&  \left. a_9 (\mathcal{D}_\rho \mathcal{D}_\lambda F^{\lambda}_\mu)^2 + \dots \right]~,
\end{eqnarray}
where $\tilde{\kappa}$ is a constant with the dimension of a squared length, $\mathcal{D}$ denotes a gauge- and gravitationally-covariant derivative, and the dots denote terms of higher powers in $\tilde{\kappa}$.

The coefficients of many of these terms may be changed by local field redefinitions
of the form~\cite{tseytlin} $A_\mu \rightarrow A_\mu = \tilde{A}_\mu + T_\mu (\tilde{A}_\mu)$, where
\begin{eqnarray}\label{redef}
&& T_\mu (A) =2 \alpha ' b_1 \mathcal{D}^\rho F_{\rho \mu} + \nonumber \\
&& (2\alpha')^2 \left[ b_2 F_{\rho\nu}\mathcal{D}_\mu F^{\rho\nu} + b_3 \mathcal{D}_\mu F^{\rho\nu}F_{\rho\nu} + \right. \nonumber \\
&& \left. b_4 \mathcal{D}_\rho F^{\rho\nu} F_{\mu\nu} + b_5 F_{\mu\nu}\mathcal{D}_\rho F^{\rho\nu} + \right. \nonumber \\
&& \left.b_6 \mathcal{D}^2 \mathcal{D}^\rho F_{\rho\mu} + \mathcal{O}(\alpha'^3)\right]~.
\end{eqnarray}
Substituting (\ref{redef}) into (\ref{efa}), one observes that the following coefficients change:
\begin{eqnarray}
a_2 & \rightarrow & a_2 - 4 b_1~, \nonumber \\
a_7 & \rightarrow & a_7 - 4 b_2 + 4b_3 - 3a_1 b_1~,  \nonumber \\
a_8 & \rightarrow & a_8 - 4 b_4 + 4 b_5 + 8b_1^2 - 6a_1 b_1 -8 a_2 b_1~, \nonumber \\
a_9 & \rightarrow & a_9 + 4 b_6 + 2b_1^2 - 2a_2 b_1 ~.
\end{eqnarray}
These coefficients are therefore ambiguous and undetermined by the scattering amplitudes. Specifically,
the $E^2$-dependent term in the effective action (\ref{greff}) arising from gravitational loop corrections corresponds to the term with coefficient $a_2$ in the effective action (\ref{efa}), and hence
does not contribute to physical processes described by on-shell scattering amplitudes~\cite{Lambdalog}.

This result is well known in string theory. Computing the three and four-point massless vector-boson scattering amplitudes using the effective Lagrangian (\ref{efa}),
comparing with the known superstring scattering amplitudes,
and identifying $\tilde{\kappa}^2 \equiv \alpha '$, the string Regge slope that plays the r\^ole of
 the gravitational scale in string theory, one finds that the coefficients $a_1=0$, $a_3=2a_4=\pi^2/3$, $a_5=2a_8=-\pi^2/12$ are fixed, while the remaining coefficients $a_2, a_7, a_8$ and $a_9$ remaining \emph{arbitrary}.
This reflects the fact that the values of the coefficients $a_1, a_3, a_4, a_5$ and $a_8$ are physically
relevant, but not the values of $a_2, a_7, a_8$ and $a_9$~\cite{EJM}.

A related point of view was taken recently in~\cite{donogh} in the context of a scalar field theory
with a $\lambda \phi^4$ interaction, where it was argued that power-law contributions to effective actions due to non-renormalizable gravitational interactions, although present, do not signify a running of the coupling
constant $\lambda$ that can be measured physically. Considering scattering amplitudes describing the
scattering of four scalar fields, the authors of \cite{donogh} demonstrated explicitly that on-shell diagrams are
finite, not suffering from any quadratic infinities. The wannabe infinities affect only higher-order operators, e.g.,
those of the form
\begin{equation}
-\lambda_1 \phi^2 \partial_\mu \phi \partial^\mu \phi~.
\end{equation}
These vanish upon using the equations of motion, and
can be removed by field redefinitions of the type $\phi \rightarrow \phi - \frac{\lambda_1}{3}\phi^3$.
In the context of our discussion above, this is another illustration of the equivalence theorem.

The above arguments apply independently of any string theory context.
However, before closing, we comment briefly on the general form of
 the unambiguous terms in the low-energy effective action of open strings propagating in background
 gauge and gravitational fields, and on the implications of our results for string unification scenarios.
 
 It is well known that the unambiguous part of the effective action includes non-linear
 Born-Infeld electrodynamics~\cite{bi}, augmented by certain specific types of higher-derivative terms
 involving the field strength~\cite{biderivative}:
\begin{eqnarray}\label{biaction}
&& S_{\rm BI} = \frac{1}{g_s} T^{p} \int d^p x \sqrt{{\rm Det}\left(g_{\mu\nu} + \alpha ' F_{\mu\nu} \right)}  \nonumber \\ &&
\times \left[ 1 + \frac{(2\pi \alpha ')^2 }{96}\left( - h^{ij} h^{k\ell} h^{mn} h^{pq} S_{npjk} S_{qm\ell i}
+ \right. \right. \nonumber \\ && \left. \left.  \frac{1}{2} h^{ij} h^{k\ell} S_{jk} S_{\ell i} + R(g) + \dots \right)\right]~,
\end{eqnarray}
where
$h^{ij} \equiv \left( \frac{1}{g + \alpha ' F}\right)^{ij}$, with $g$ denoting the background space-time metric, is an effective open string (inverse) metric,
the indices $i,j,k,\ell$ are space-time indices running over the appropriate longitudinal dimensions
of the brane hypersurfaces to which open strings are attached, $T$ is the string tension, $g_s$ is the string coupling, $\alpha'$ is the square of the string length, and the quantities :
\begin{eqnarray}
&& S_{npjk} \equiv \partial_n \partial_p F_{jk} + 2 (2\pi \alpha ') h^{rs} \partial_n F_{jr} \partial_p F_{ks}~, \nonumber \\ && S_{jk} \equiv h^{mn}S_{jkmn}~.
\end{eqnarray}
The $\dots$ in (\ref{biaction}) represent higher-derivative contributions, and the terms $R(g)$ arise from curvature of the brane worlds.

We notice that six-derivative terms that are bilinear in $A_\mu$ can be identified in the above action by
setting $h^{ij} = g^{ij} $, which implies that the corresponding $S_{jk} =0$ (because of the antisymmetry of $F_{jk}=-F_{kj}$). It is then straightforward to see that these terms acquire the form
$F_{jk} (\Box)^2 F^{jk}$ (up to a coefficient). According to our discussion above, this can be removed by a field redefinition, leaving only interaction terms with higher-order couplings as unambiguous, in the sense of
contributing to on-shell string scattering amplitudes.

It follows that, when comparing gauge and gravitational interaction strengths in string unification scenarios,
there are no relevant power-law gravitational loop corrections.

In four space-time dimensions, corresponding to open strings attached to three-branes,
the Born-Infeld determinant can be written as~\cite{bi}:
\begin{eqnarray}
&&{\rm Det}_4 = g \left( 1 + 2I_2 - I_4\right): \nonumber \\
&& I_2 = \frac{1}{4}F_{\mu\nu}^2~, \nonumber \\
&& I_4 = \frac{1}{16}(F_{\mu\nu}\,F^{*\mu\nu})^2~,
\end{eqnarray}
where $F^{*\mu\nu} = \frac{1}{2}\epsilon^{\mu\nu\rho\sigma}F_{\rho\sigma}$.
Expanding the square root in (\ref{biaction}) space-time derivatives, and recalling that
the string tension in this case is just $T=1/(2\pi\alpha')$, we
make direct contact with four-dimensional QED to lowest order in derivatives in the presence of a cosmological constant at order one in units of $\alpha'$, arising from the metric $g$ term in the action (\ref{biaction}). In such a case, $g_s$ is identified with the square of the electric charge $e^2$.
Any string-loop gravitational corrections to this action are computed by considering open-string $\sigma$ models on  two-dimensional world-sheet surfaces with higher genus. In the low-energy limit, any effective gravitational corrections to terms quadratic in $F_{\mu\nu}^2$ and in gauge and gravitational covariant derivatives would be absorbed by tree-level field redefinitions, as described above, and thus would not contribute to the scattering amplitudes or to physical running couplings.

In the presence of a dilaton field, there is an overall factor $e^{-\phi}$ multiplying the Lagrangian (\ref{biaction}), in addition to terms with derivatives of the dilaton. In non-constant time-dependent dilaton backgrounds, the cosmological constant may relax~\cite{ABEN}, 
and in such a case one may also have time-dependent gauge couplings.
Perturbative $S$-matrix elements may be well defined in some relaxation models for the cosmological constant, and in such cases our considerations on the absorption of $E^2$ terms in local field redefinitions would apply intact.
On the other hand, in the cases of a fixed positive cosmological constant and some quintessence models
(which could be obtained in string theory via one-string-loop dilaton tadpoles~\cite{tadpole}) the $S$-matrix is not well defined~\cite{smatrix}. In such a case the usual equivalence theorem would not apply. 
%However, the corresponding $\Lambda$-dependent gravitational loop corrections to the effective action depend only logarithmically on the energy scale, and hence may contribute to the conventional running of the gauge coupling.

We conclude by restating our principal conclusion, which is independent of the string theory context:
the equivalence theorem~\cite{equiv} implies that $S$-matrix elements are unaffected by higher-order
derivative corrections to terms in the effective gauge theory action that are quadratic in the gauge fields.

{\bf Acknowledgements:}
We thank Luis \'Alvarez-Gaum\'e for discussions.
One of (NEM) thanks the CERN Theory Division for its kind hospitality.


\begin{thebibliography}{99}


\bibitem{wilczek} S.~P.~Robinson and F.~Wilczek,
  %``Gravitational correction to running of gauge couplings,''
  Phys.\ Rev.\ Lett.\  {\bf 96}, 231601 (2006)
  [arXiv:hep-th/0509050].
  %%CITATION = PRLTA,96,231601;%%

\bibitem{gauge} A.~R.~Pietrykowski,
  %``Gauge dependence of gravitational correction to running of gauge
  %couplings,''
  Phys.\ Rev.\ Lett.\  {\bf 98}, 061801 (2007)
  [arXiv:hep-th/0606208].
  %%CITATION = PRLTA,98,061801;%%



\bibitem{absent} D.~J.~Toms,
  %``Quantum gravity and charge renormalization,''
  Phys.\ Rev.\  D {\bf 76}, 045015 (2007)
  [arXiv:0708.2990 [hep-th]].
  %%CITATION = PHRVA,D76,045015;%%


\bibitem{various} D.~Ebert, J.~Plefka and A.~Rodigast,
  %``Absence of gravitational contributions to the running Yang-Mills
  %coupling,''
  Phys.\ Lett.\  B {\bf 660}, 579 (2008)
  [arXiv:0710.1002 [hep-th]].
  %%CITATION = PHLTA,B660,579;%%

\bibitem{other} Y.~Tang and Y.~L.~Wu, Comm. Theor. Phys. \textbf{54}, 1040 (2010),
  %``Gravitational Contributions to the Running of Gauge Couplings,''
  arXiv:0807.0331 [hep-ph].
  %%CITATION = NONE,,;%%



\bibitem{toms} D.~J.~Toms,
  %``Quantum gravitational contributions to quantum electrodynamics,''
  Nature {\bf 468}, 56 (2010)
  [arXiv:1010.0793 [hep-th]].
  %%CITATION = NATUA,468,56;%%

\bibitem{recent} Y.~Tang and Y.~L.~Wu,
  %``Quantum Gravitational Contributions to Gauge Field Theories,''
  arXiv:1012.0626 [hep-ph].
  %%CITATION = ARXIV:1012.0626;%%

\bibitem{equiv}
A.~Salam and J.~A.~Strathdee,
  %``Equivalent formulations of massive vector field theories,''
  Phys.\ Rev.\  D {\bf 2}, 2869 (1970);
  %%CITATION = PHRVA,D2,2869;%%
R.E. Kallosh and I.V. Tyupin, Yad. Fiz. \textbf{17}, 190 (1970);
M.C. Berger and Y.M.P. Lam, Phys. Rev. D \textbf{13}, 3247 (1976).
See also S.~Weinberg,
  %``Nonlinear realizations of chiral symmetry,''
  Phys.\ Rev.\  {\bf 166}, 1568 (1968).
  %%CITATION = PHRVA,166,1568;%%

\bibitem{tseytlin} A.~A.~Tseytlin,
  %``Ambiguity in the Effective Action in String Theories,''
  Phys.\ Lett.\  B {\bf 176}, 92 (1986).
  %%CITATION = PHLTA,B176,92;%%

\bibitem{fermions}
In the presence of fermions, such redefinitions would yield higher-derivative corrections in the fermionic
sector of the effective Lagrangian, such as terms of the form
$\frac{c}{M_P^2} \int d^4 x \overline{\psi} \gamma^\mu \nabla^\nu F_{\mu\nu} \psi$.

\bibitem{Lambdalog}
This discussion does not extend to the cosmological constant term in (\ref{greff}), which depends
logarithmically on the energy scale, and hence cannot be absorbed by a local field redefinition.
On the other hand, in the presence of a positive cosmological constant $\Lambda$ perturbative
$S$-matrix elements cannot be defined, requiring a different discussion, see below.

\bibitem{EJM}
We recall also that an explicit one-loop heterotic string calculation demonstrated that the
Chamseddine-Chapline-Manton action is unrenormalized, whereas terms of higher order in the gauge
fields are generated:
J.~R.~Ellis, P.~Jetzer and L.~Mizrachi,
  %``ONE LOOP STRING CORRECTIONS TO THE EFFECTIVE FIELD THEORY,''
  Nucl.\ Phys.\  B {\bf 303}, 1 (1988),
  %%CITATION = NUPHA,B303,1;%%
 and that this argument can be extended to multiple loops:
 J.~R.~Ellis and L.~Mizrachi,
  %``MULTILOOP NO RENORMALIZATION THEOREMS FOR THE TEN-DIMENSIONAL HETEROTIC
  %STRING,''
  Nucl.\ Phys.\  B {\bf 327}, 595 (1989).
  %%CITATION = NUPHA,B327,595;%%

\bibitem{donogh} M.~M.~Anber, J.~F.~Donoghue and M.~El-Houssieny,
  %``Running couplings and operator mixing in the gravitational corrections to
  %coupling constants,''
  arXiv:1011.3229 [hep-th].
  %%CITATION = ARXIV:1011.3229;%%

\bibitem{bi} E.~S.~Fradkin and A.~A.~Tseytlin,
  %``Nonlinear Electrodynamics From Quantized Strings,''
  Phys.\ Lett.\  B {\bf 163}, 123 (1985);
  %%CITATION = PHLTA,B163,123;%%
R.~R.~Metsaev, M.~Rakhmanov and A.~A.~Tseytlin,
  %``THE BORN-INFELD ACTION AS THE EFFECTIVE ACTION IN THE OPEN SUPERSTRING
  %THEORY,''
  Phys.\ Lett.\  B {\bf 193}, 207 (1987);
  %%CITATION = PHLTA,B193,207;%%
For a modern comprehensive review and properties of Born-Infeld action in superstring theory see:
A.~A.~Tseytlin,
  %``Born-Infeld action, supersymmetry and string theory,''
  arXiv:hep-th/9908105.
  %%CITATION = HEP-TH/9908105;%%

\bibitem{biderivative} N.~Wyllard,
  %``Derivative corrections to D-brane actions with constant background
  %fields,''
  Nucl.\ Phys.\  B {\bf 598}, 247 (2001)
  [arXiv:hep-th/0008125].
  %%CITATION = NUPHA,B598,247;%%

\bibitem{ABEN}
I.~Antoniadis, C.~Bachas, J.~R.~Ellis and D.~V.~Nanopoulos,
  %``Cosmological String Theories and Discrete Inflation,''
  Phys.\ Lett.\  B {\bf 211}, 393 (1988) and
  %%CITATION = PHLTA,B211,393;%%
%``An Expanding Universe in String Theory,''
  Nucl.\ Phys.\  B {\bf 328}, 117 (1989).
  %%CITATION = NUPHA,B328,117;%%

\bibitem{tadpole} W.~Fischler and L.~Susskind,
  %``Dilaton Tadpoles, String Condensates and Scale Invariance,''
  Phys.\ Lett.\  {\bf B171}, 383 (1986);
%``Dilaton Tadpoles, String Condensates and Scale Invariance. 2.,''
 \emph{ibid}.  {\bf B173}, 262 (1986).

\bibitem{smatrix}  See, for instance: S.~Hellerman, N.~Kaloper and L.~Susskind,
  %``String theory and quintessence,''
  JHEP {\bf 0106}, 003 (2001).
  [hep-th/0104180];
W.~Fischler, A.~Kashani-Poor, R.~McNees and S.~Paban,
  %``The Acceleration of the universe, a challenge for string theory,''
  JHEP {\bf 0107}, 003 (2001).
  [hep-th/0104181]. 

\end{thebibliography}
\end{document}